\documentstyle[12pt]{article}
\begin{document}
\vspace{10mm}
\begin{center}
{\large \bf RECENT DEVELOPMENTS IN LOW $x$ PHYSICS \footnote {Based 
on lectures given 
at the XXXVIth Cracow School of Theoretical Physics, 
Zakopane, Poland, June 1996}}\\
\vspace{8mm}
J. Kwieci\'nski \\
Department of Theoretical Physics\\
H. Niewodnicza\'nski Institute of Nuclear Physics\\
Krak\'ow, Poland. \\
\end{center}
\vspace{20mm}
\begin{abstract}
The QCD expectations for the 
behaviour of the deep inelastic scattering structure functions in the 
region of small values of the Bjorken parameter $x$ are summarized.   
 The Balitzkij, Lipatov, 
Fadin, Kuraev (BFKL) equation which sums the leading powers of 
$\alpha_s ln(1/x)$ is described and confronted with the "conventional" 
formalism based on the 
leading  order QCD evolution equations. The small $x$ behaviour of the spin dependent 
structure function $g_1$ is also discussed. The dedicated measurements of the hadronic final
state in deep inelastic scattering at small $x$ probing the QCD 
pomeron are described.  This includes discussion of the 
deep inelastic scattering accompanied by forward jets as well as production 
of forward prompt photons and $\pi^0$-s.  Finally the basic facts concerning the 
deep inelastic diffraction are briefly summarized.    
\end{abstract}
\medskip\medskip
\section*{1. Introduction}
The advent of the HERA $ep$ collider has opened up a possibility to test 
QCD in the new and hitherto unexplored regime of the small values of the 
Bjorken parameter $x$.  This parameter  
is, as usual, defined  as $x=Q^2/(2pq)$ where $p$ is the proton four momentum, 
$q$ 
the four momentum transfer between the leptons and $Q^2=-q^2$. 
  Perturbative QCD 
predicts that several new phenomena will occur when the parameter 
$x$ specifying 
the longitudinal momentum fraction of a hadron carried by a parton 
(i.e. by a quark or by a gluon) 
becomes very small \cite{GLR,BCKK} . The main expectation is
  that the gluon  densities 
should  strongly grow in this limit, eventually leading to the parton 
saturation effects \cite{GLR,BCKK,ADM1,JK1}. 
  The small $x$ behaviour of the structure functions 
is driven by the gluon through the $g \rightarrow q \bar q$ transition  
and the  increase of gluon distributions with decreasing $x$  
implies a similar increase of the deep inelastic lepton - proton  
 scattering structure function $F_2$ 
as the  Bjorken parameter $x$ decreases \cite{AKMS}.  
The recent experimental data are consistent with 
this perturbative QCD prediction that the structure function $F_2(x,Q^2)$ 
should strongly grow with the decreasing Bjorken parameter 
$x$ \cite{H1,ZEUS,BBDUR,LANC,HALINA}.\\      

The purpose of these lectures is to summarize the QCD expectations for the small $x$ 
behaviour of the deep inelastic scattering structure functions.  After briefly 
reviewing predictions of the Regge theory we shall discuss the Balitzkij, Lipatov, 
Fadin, Kuraev (BFKL) \cite{BFKL,LIPATOV} equation which sums the leading powers of 
$\alpha_s ln(1/x)$.  
We will also briefly describe the "conventional" formalism based on the 
leading (and next to leading) order QCD evolution equations and confront it 
with the BFKL equation.  Besides the  structure functions 
$F_2(x,Q^2)$ and $F_L(x,Q^2)$ we shall also consider the spin  
structure function $g_1(x,Q^2)$.  The novel feature in the latter case is  
 the appearence of the double logarithmic terms i.e. powers of 
$\alpha_sln^2(1/x)$ at each order of the perturbative expansion. Finally we 
will also discuss dedicated measurements of the final state in deep 
inelastic scattering which can probe the detailed dynamical 
content of the QCD pomeron. \\
 
\section*{2. Structure functions at low $x$}
Small $x$ behaviour of structure functions  
for fixed $Q^2$ reflects the high energy behaviour of the 
virtual Compton scattering total cross-section with increasing   
total CM energy 
squared $W^2$ since $W^2=Q^2(1/x-1)$. The appropriate framework 
for the theoretical description 
of this behaviour is the Regge pole exchange picture  
\cite{PC} .\\

The high energy behaviour of the total 
hadronic and (real) photoproduction cross-sections can be economically 
described by two contributions: an (effective) pomeron with its 
intercept slightly above unity ($\sim 1.08$) and the leading meson 
Regge trajectories  
with  intercept $\alpha_R(0) \approx 0.5$ \cite{DOLA}.  
The reggeons can be 
identified as corresponding to $\rho, \omega$, $f$ or $A_2$ 
exchange(s) depending 
upon the quantum numbers involved. All these 
reggeons have approximately the same intercept.   
One refers to the pomeron obtained 
from the phenomenological analysis of  hadronic  total cross 
sections as the "soft" pomeron since the bulk of the processes building-up 
the cross sections are  low $p_t$ (soft) processes.\\

The Regge pole model gives the following parametrization of the deep 
inelastic scattering structure function $F_2(x,Q^2)$ at  
small $x$ \cite{DOLAF2,IOFFE}
\begin{equation}
F_2(x,Q^2)=\sum_i \tilde \beta_i(Q^2) x^{1-\alpha_i(0)}. 
\label{reggef}
\end{equation}
The relevant reggeons are 
those which can couple to two (virtual) photons.  The (singlet) part 
of the structure function $F_2$ is controlled at small $x$ by 
pomeron exchange, while the non-singlet part $F_2^{NS}=
F_2^p-F_2^n$ by the  $A_2$ reggeon.  
Neither pomeron nor $A_2$ reggeons  couple to the 
spin structure function $g_1(x,Q^2)$ which is described at small 
$x$ by the  exchange of reggeons corresponding to 
axial vector mesons \cite{IOFFE,EKARL} i.e. to  $A_1$ exchange for the non-singlet 
part $g_1^{NS} = g_1^{p}-g_1^{n}$ etc.
\begin{equation}
g_1^{NS}(x,Q^2)=\gamma(Q^2) x^{-\alpha_{A_1}(0)}. 
\label{gnsa1}
\end{equation}
  The  reggeons which correspond to axial vector mesons 
are expected to 
have very low intercept (i.e. $\alpha_{A_1} \le 0$ etc.).\\

The experimental results from HERA show  that 
the proton structure function $F_{2}(x,Q^2)$ for moderate and large $Q^2$ values 
($Q^2 >$ 1.5 $GeV^2$ or so) grows more rapidly than expected on the basis of the 
straightforward extension of the Regge pole parametrization with the 
relatively small intercept of the effective pomeron ($\alpha_P(0) 
\approx$ 1.08) {\cite{HALINA}. This result is consistent with  
 perturbative QCD which predicts much stronger increase of the parton 
 distributions and of the DIS structure functions with decreasing parameter 
 $x$ than that which would follow from equation (\ref{reggef}) with 
 $\alpha_P(0) \approx$ 1.08.  The high energy behaviour which follows 
 from perturbative QCD is often referred to as being related to 
 the "hard" 
 pomeron in contrast to the soft pomeron describing the high energy 
 behaviour of hadronic and photoproduction cross-sections.\\

 The relevant framework for discussing the pomeron in
perturbative QCD and the small $x$ limit of parton 
distributions is the leading $ln 1/x$ (LL$1/x$) approximation which 
corresponds to the sum of those terms in the perturbative expansion 
where the powers of $\alpha_s$ are accompanied by the leading 
powers of ln($1/x$) \cite{GLR,BCKK,ADM1,JK1,BFKL,LIPATOV}.   
At small $x$ the dominant role is played by the 
gluons and the quark (antiquark) distributions as well as the deep 
inelastic structure functions $F_{2,L}(x,Q^2)$ are also driven by the 
gluons through the $g \rightarrow q\bar q$ transitions.
Dominance of gluons in high energy scattering follows from the
fact that they carry spin equal to unity. 
The  basic dynamical quantity at small $x$ is the   
unintegrated gluon distribution 
$f(x,Q_t^2)$ where $x$ denotes the momentum fraction 
of a parent hadron carried by a gluon and $Q_t$  its transverse 
momentum.  The unintegrated distribution $f(x,Q_t^2)$ 
is related in the following way to the more familiar scale dependent 
gluon distribution $g(x,Q^2)$: 
\begin{equation}
xg(x,Q^2)=\int^{Q^2} {dQ_t^2\over Q_t^2} f(x,Q_t^2). 
\label{intg}
\end{equation}
In the leading $ln(1/x)$  approximation the unintegrated 
distribution $f(x,Q_t^2)$ satisfies 
the BFKL equation \cite{BFKL,LIPATOV,CIAF} which has the following form: 
$$
f(x,Q_t^2)=f^0(x,Q_t^2)+
$$
\begin{equation}
\bar \alpha_s \int_x^1{dx^{\prime}\over 
x^{\prime}} \int {d^2 q\over \pi q^2}
\left[{Q_t^2 \over (\mbox{\boldmath $q$}+
\mbox{\boldmath $Q_t$})^2} 
f(x^{\prime},(\mbox{\boldmath $q$}+
\mbox{\boldmath $Q_t$})^2)-f(x^{\prime},Q_t^2)\Theta(Q_t^2-q^2)\right]
\label{bfkl}
\end{equation}
where 
\begin{equation}
\bar \alpha_s={3\alpha_s\over \pi}
\label{alphab}
\end{equation}
This equation sums the ladder diagrams with gluon exchange accompanied 
by virtual corrections which are responsible for the gluon reggeization. 
The first and the second 
terms  on the right hand side of  eq. (\ref{bfkl}) correspond 
to  real gluon emission with $q$ being the transverse 
momentum of the emitted gluon, and to the virtual corrections respectively. 
$f^0(x,Q_t^2)$ is a suitably defined inhomogeneous term.\\
 
For the fixed coupling case  eq. (\ref{bfkl}) can be solved 
analytically and the leading behaviour of its solution 
at small $x$ is given by the 
following expression:
\begin{equation} 
f(x,Q_t^2) \sim (Q_t^2)^{{1\over 2}} {x^{-\lambda_{BFKL}}\over 
\sqrt{ln({1\over x})}} exp\left(-{ln^2(Q_t^2/\bar Q^2)\over 2 \lambda^"
ln(1/x)} \right)
\label{bfkls}
\end{equation} 
with 
\begin{equation}
\lambda_{BFKL}=4 ln(2) \bar \alpha_s
\label{pombfkl}
\end{equation} 
\begin{equation} 
\lambda^"=\bar \alpha_s 28 \zeta(3) 
\label{diff}
\end{equation}
where the Riemann zeta function $\zeta(3) \approx 1.202$.  The 
parameter $\bar Q$ is of nonperturbative origin.\\

The quantity $1+ \lambda_{BFKL}$ is equal to the intercept of the so -  
called BFKL pomeron. Its potentially large magnitude ($\sim 1.5$) 
should be contrasted with the intercept $\alpha_{soft} \approx 1.08$ 
of the (effective) "soft" pomeron which has been determined 
from the phenomenological analysis of the high energy behaviour 
of hadronic and photoproduction total cross-sections \cite{DOLA}.\\   
  
The solution of the BFKL equation 
reflects its diffusion pattern which  
is the direct consequence of the absence of transverse momentum ordering 
along the gluon chain.   
The interrelation between the diffusion of transverse momenta towards 
both the infrared and ultraviolet regions {\bf and} the increase of gluon 
distributions 
with decreasing $x$ is a  characteristic property of QCD at low $x$.  
It has important consequences for the structure of the hadronic final 
state in deep inelastic scattering at small $x$ and this problem will be 
discussed with some detail in the next Section. \\

In practice one introduces the running coupling $\bar \alpha_s(Q_t^2)$ 
in the BFKL equation (\ref{bfkl}). This requires the introduction of an 
infrared 
cut-off to prevent entering the infrared region where the 
coupling becomes large. The effective intercept $\lambda_{BFKL}$ 
found by numerically solving the equation depends   
on the magnitude of this cut-off \cite{KMS2}. The running coupling does 
also affect the diffusion pattern of the solution. 
 The effective intercept $\lambda_{BFKL}$ turns
 out 
to be also sensitive 
on the (formally non-leading) additional constraint $q^2 < Q_t^2x^{\prime}/x$ 
in the real emission term in eq. (\ref{bfkl}) which 
follows from the requirement that the virtuality of the 
last gluon in the chain is dominated by $Q_t^2$ \cite{BO,KMSGLU}. 
For fixed coupling $\bar \alpha_s$ the modified BFKL equation with the 
constraint 
$q^2 < Q_t^2x^{\prime}/x$ can be solved analytically since imposing this 
constraint still preserves the scale invariance of this equation.  One can also 
derive the NLO formula for  $\lambda_{BFKL}$ which reads:
\begin{equation}
\lambda_{BFKL} =\bar \alpha_s 4ln 2 (1 - 4.15 \bar \alpha_s)
\label{nlo}
\end{equation}
In Fig. 1 we show  $\lambda_{BFKL}$ plotted as a function of  
$\bar \alpha_s$ We also show the LO (\ref{pombfkl}) and NLO  (\ref{nlo}) 
results.  We see that the constraint $q^2 < Q_t^2x^{\prime}/x$ significantly 
affects the magnitude of  $\lambda_{BFKL}$.  It may also be seen that 
the pure NLO result may be unreliable even for values of $\bar \alpha_s \sim $
 0.1. \\
   
The impact of the momentum 
cut-offs on the solution of the BFKL equation has also been discussed 
in refs. \cite{PVLC,MCDG}. In  impact parameter representation 
the BFKL equation offers an 
interesting  
interpretation in terms of colour dipoles \cite{DIPOLE}.  It 
should also be emphasised that the complete calculation of the next-to-leading 
corrections to the BFKL equation has recently become presented in ref.
 \cite{NLX}.\\

The structure functions $F_{2,L}(x,Q^2)$ are driven  at small $x$ 
by the gluons 
 and are related in the following way to the unintegrated distribution $f$: 
\begin{equation}
F_{2,L}(x,Q^2)=\int_x^1{dx^{\prime}\over x^{\prime}}\int 
{dQ_t^2\over Q_t^2}F^{box}_{2,L}(
x^{\prime},Q_t^2,Q^2)f({x\over x^{\prime}},Q_t^2). 
\label{ktfac}
\end{equation}
The functions  $F^{box}_{2,L}(x^{\prime},Q_t^2,Q^2)$ may be regarded as  the 
structure 
functions of the off-shell gluons with  virtuality  
$Q_t^2$.  
They are described by the quark box (and crossed box) diagram contributions 
to the 
photon-gluon interaction.   
The small $x$ behaviour of the structure functions reflects the small 
$z$ ($z = x/x^{\prime}$) behaviour of the gluon distribution $f(z,Q_t^2)$.\\

Equation (\ref{ktfac}) is an example of the "$k_t$ factorization theorem" 
which relates measurable quantities (like DIS structure functions) to 
the convolution in both longitudinal as well as in transverse momenta of the 
universal gluon distribution $f(z,Q_t^2)$ with the cross-section 
(or structure function) describing the interaction of the "off-shell" gluon 
with the hard probe \cite{KTFAC,CIAFKT,MK}.  
The $k_t$ factorization theorem is the basic tool for 
calculating the observable quantities in the small $x$ region in terms of the 
(unintegrated) gluon distribution $f$ which is the solution of the BFKL 
equation.\\ 

The leading - twist part of the $k_t$ factorization formula can be rewritten 
in a collinear factorization form.  The leading small $x$ effects are then 
automatically resummed in the  
 splitting functions and in the coefficient functions. The $k_t$ 
factorization theorem   can in fact be used as the tool for calculating 
these quantities.   Thus, for instance, the moment function $\bar 
P_{qg}(\omega, 
\alpha_s)$ of the splitting $P_{qg}(z, 
\alpha_s)$ function is represented in the following form 
(in the so called $Q_0^2$ regularization and DIS scheme \cite{CIAFKT}): 

\begin{equation}
\bar P_{qg}(\omega,\alpha_s)= 
{\gamma_{gg}^{2}({\bar \alpha_s\over \omega})\tilde F^{box}_{2}
\left(\omega=0,\gamma= \gamma_{gg}({\bar \alpha_s\over \omega})\right)
\over 2\sum_i e_i^2 }
\label{pqgf}
\end{equation} 
where  $\tilde F^{box}_{2}(
\omega,\gamma)$ is the Mellin transform of the moment function 
$\bar F^{box}_{2}(
\omega, Q_t^2,Q^2)$ i.e. 
\begin{equation}
\bar F^{box}_{2}(
\omega, Q_t^2,Q^2)={1\over 2 \pi i} \int_{1/2-i\infty}^{1/2+i\infty} 
d\gamma \tilde F^{box}_{2}(
\omega,\gamma)\left(Q^2\over Q_t^2 \right)^{\gamma}
\label{mbox}
\end{equation}
and the anomalous dimension $\gamma_{gg}({\bar \alpha_s\over \omega})$ 
has the following expansion \cite{JAR}; 
\begin{equation}
\gamma_{gg}({\bar \alpha_s\over \omega})=\sum_{n=1}^{\infty}c_n
\left({\bar \alpha_s\over \omega}\right)^n
\label{adexp}
\end{equation}
This expansion gives the following expansion of the splitting function 
$P_{gg} $
\begin{equation}
zP_{gg}(z,\alpha_s)=\sum_1^{\infty}c_n{[\alpha_s ln(1/z)]^{n-1}\over (n-1)!}
\label{pggzet}
\end{equation}
Representation (\ref{pqgf}) generates the following  
expansion of the splitting function $P_{qg}(z,\alpha_s)$ at small $z$:  
\begin{equation}
zP_{qg}(z,\alpha_s)={\alpha_s\over 2 \pi}zP^{(0)}(z) +
(\bar \alpha_s)^2 \sum_{n=1}^{\infty}b_n
{[\bar \alpha_s ln(1/z)]^{n-1}\over (n-1)!}
\label{zpqgf}
\end{equation}
  The first term on the right hand side of eq. (\ref{zpqgf})
vanishes at $z=0$.  It should be noted that 
 the  splitting function $P_{qg}$ 
is formally non-leading at small $z$ when compared with the splitting  
function $P_{gg}$ .   
For moderately small values of $z$ however,  
when the first few terms in the expansions (\ref{adexp}) and (\ref{zpqgf})
dominate, the BFKL effects can be much more important 
in $P_{qg}$  than in $P_{gg}$.  
This comes from the fact that in the expansion (\ref{zpqgf}) 
all coefficients $b_n$ are different from zero while in eq. (\ref{adexp}) 
we have $c_2=c_3=0$ \cite{JAR}.  
The small $x$ resummation effects within the conventional QCD evolution 
formalism have recently been discussed in refs. 
\cite{EKL,HBRW,BFORTE,FRT,CIAFQQ,CATHEP}. 
  One finds in general that at the moderately small 
values of $x$ which are relevant for 
the HERA measurements,   the small $x$ resummation effects in the 
splitting function $P_{qg}$ have a much stronger impact on $F_{2}$ than 
the small $x$ resummation in the splitting function $P_{gg}$. This  
reflects  the fact, which has already been mentioned 
above, that  in the expansion (\ref{zpqgf}) 
all coefficients $b_n$ are different from zero while in eq. (\ref{adexp}) 
we have $c_2=c_3=0$. It should also be remembered that the BFKL effects 
in the splitting function $P_{qg}(z,\alpha_s)$ can significantly affect 
extraction of the gluon distribution out of the experimental data on the 
slope of the structure function $F_2(x,Q^2)$ which is based on the following relation:
\begin{equation}
Q^2{\partial F_2(x,Q^2)\over \partial Q^2} \simeq 
2 \sum_i e_i^2 \int_x^1 dz P_{qg}(z,\alpha_s(Q^2)){x\over z}
g({x\over z},Q^2)
\label{slopef2}
\end{equation}
 
A more general treatment of the gluon ladder than that which follows 
from the BFKL formalism is  provided by 
the Catani, Ciafaloni, Fiorani, Marchesini (CCFM) equation based on
 angular ordering along the gluon chain 
\cite{CCFM,KMS1,MTK}.  
This equation embodies both the BFKL equation at small $x$ and the 
conventional Altarelli-Parisi evolution at large $x$.  
The unintegrated gluon distribution $f$ now acquires   
dependence upon an additional scale $Q$ 
which specifies the maximal angle of gluon 
emission.  
The CCFM equation has the following form : 
$$
f(x,Q_t^2,Q^2)=\hat f^0(x,Q_t^2,Q^2)+ $$
\begin{equation}
\bar \alpha_s
\int_x^1{dx^{\prime}\over 
x^{\prime}} \int {d^2 q\over \pi q^2} \Theta
(Q-qx/x^{\prime})\Delta_R({x\over x^{\prime}},Q_t^2,q^2)
{Q_t^2 \over (\mbox{\boldmath $q$}+
\mbox{\boldmath $Q_t$})^2} 
f(x^{\prime},(\mbox{\boldmath $q$}+
\mbox{\boldmath $Q_t$})^2,q^2))      
\label{ccfm}
\end{equation}
where the theta function $\Theta(Q-qx/x^{\prime})$ reflects the angular 
ordering constraint on the emitted gluon.  
The "non-Sudakov" form-factor $\Delta_R
(z,Q_t^2,q^2  )$ is now given by the following formula: 
\begin{equation}
\Delta_R(z,Q_t^2,q^2)=exp\left[-\bar \alpha_s\int_z^1 {dz^{\prime}
\over z^{\prime}} \int {dq^{\prime 2}
\over q^{\prime 2}}\Theta (q^{\prime 2}-(qz^{\prime})^2)
\Theta (Q_t^2-q^{\prime 2})\right]
\label{ns}
\end{equation}
Eq.(\ref{ccfm}) still contains only the singular term of the 
$g \rightarrow gg$ splitting function  
at small $z$. Its generalization which would  
include 
remaining parts of this vertex (as well as quarks) is possible.  
The numerical analysis of this equation was presented in ref. \cite{KMS1} 
. The CCFM equation which is the generalization of the BFKL equation generates 
the steep $x^{-\lambda}$ type of behaviour for the deep inelastic structure 
functions as the effect of the leading $ln(1/x)$ 
resummation \cite{CCFMF2}.  The slope $\lambda$ turns out to be sensitive 
on the (formally non-leading) additional constraint $q^2 < Q_t^2x^{\prime}/x$ 
in eq. (\ref{ccfm}) which 
follows from the requirement that the virtuality of the 
last gluon in the chain is dominated by $Q_t^2$ \cite{BO,KMSGLU}.\\

The HERA data can be described quite well using the BFKL and CCFM equations 
combined with the factorization formula (\ref{ktfac})  
\cite{KMSGLU,CCFMF2}. One can however obtain satisfactory description of the HERA data staying 
within the scheme   
 based on the Altarelli-Parisi equations alone 
without the small $x$ resummation effects being included in the formalism 
\cite{MRS,GRV,MRSR}.   
In the latter case the singular small $x$ behaviour of the gluon 
and sea quark distributions  
 has to be introduced in the parametrization of the starting 
distributions at the moderately large reference scale $Q^2=Q_0^2$   
 (i.e. $Q_0^2 \approx 4 GeV^2$ or so) \cite{MRS}.  One can also 
generate steep behaviour dynamically starting from  
non-singular "valence-like" parton distributions at some very low 
scale $Q_0^2=0.35GeV^2$ \cite{GRV}. In the latter case the gluon and sea 
quark 
distributions exhibit  "double logarithmic behaviour" \cite{DL} 
\begin{equation}
F_2(x,Q^2) \sim exp \left(2\sqrt{\xi(Q^2,Q_0^2)ln(1/x)}\right)
\label{dlog}
\end{equation}
where 
\begin{equation}
\xi(Q^2,Q_0^2)=\int_{Q_0^2}^{Q^2}{dq^2\over q^2}{3\alpha_s(q^2)\over \pi} . 
\label{evlength}
\end{equation} 
For very small values of the scale $Q_0^2$ the evolution length $\xi(Q^2,
Q_0^2)$  
can become large for moderate and large values of $Q^2$ and the "double 
logarithmic" behaviour (\ref{dlog}) is, within the limited region of $x$,  
similar to that corresponding to the power like increase of the type 
$x^{-\lambda}$, $\lambda \approx 0.3$.  In Fig. 2 we summarize results  
of the recent QCD parton model analysis of the low $x$ data from HERA 
and from the E665 experiment at Fermilab \cite{MRSR}. \\ 

The discussion presented above concerned the small $x$ 
behaviour of the singlet structure function which was driven by the gluon 
through the $g \rightarrow q \bar q$ transition. 
The gluons of course decouple from the non-singlet channel and the 
mechanism of generating the small $x$ behaviour in this case is different.\\

The novel feature of the non-singlet channel is the appearence of the 
{\bf double} logarithmic  terms i.e. powers of 
$\alpha_s ln^2(1/x)$ at each order of the perturbative 
expansion \cite{GORSHKOV,JK2,KL,EMR,BER}.  
These double logarithmic terms are generated by the 
ladder diagrams with  quark (antiquark) exchange along the chain.   
The ladder diagrams can acquire corrections from the "bremsstrahlung" 
contributions \cite{KL,BER}  
which do not vanish for the polarized structure function 
$g_1^{NS}(x,Q^2)$ \cite{BER}. They are  however relatively unimportant and 
are non-leading in the $1/N_c$ expansion. \\

In the approximation where the leading double logarithmic terms 
are generated by ladder diagrams with quark (antiquark) exchange along 
the chain  the 
unintegrated non-singlet spin dependent quark distribution $\Delta 
f_q^{NS}(x,k^2_t)$ 
($\Delta f_q^{NS}=\Delta u+\Delta \bar u - \Delta d -\bar \Delta d$) satisfies 
the following integral equation  :     
\begin{equation}
\Delta f_q^{NS}(x,Q_t^2)=\Delta f_{q0}^{NS}(x,Q_t^2)+ \tilde \alpha_s 
\int_x^1{dz\over z}\int_{Q_0^2}^{{Q_t^2\over z}} {dQ_t^{\prime 2}\over 
Q_t^{\prime 2}}\Delta f_q^{NS}({x\over z},Q_t^{\prime 2})
\label{dleq}
\end{equation}
where 
\begin{equation}
\tilde \alpha_s = {2 \over 3 \pi} \alpha_s 
\label{atil}
\end{equation} 
and $Q_0^2$ is the infrared cut-off parameter. 
The upper limit $Q_t^2/z$ in the integral equation (\ref{dleq}) follows 
from the 
requirement that the virtuality of the quark at the end of the chain 
is dominated by $Q_t^2$. A possible non-perturbative $A_1$ reggeon contribution 
has to be introduced in the driving term i.e. 
\begin{equation}
\Delta f_{q0}^{NS}(x,Q_t^2) \sim x^{-\alpha_{A_1}(0)}
\label{a2driv}
\end{equation}
at small $x$.\\
  
Equation (\ref{dleq}) implies the following equation 
for the moment function $\Delta \bar f_q^{NS}(\omega,Q_t^2)$ 
\begin{equation}
\Delta \bar f_q^{NS}(\omega,Q_t^2)=\Delta \bar f_{q0}^{NS}(\omega,Q_t^2)+ 
{\tilde \alpha_s \over \omega} \left[\int_{Q_0^2}^{Q_t^2} 
{dQ_t^{\prime 2}\over 
Q_t^{\prime 2}}\Delta \bar f_q^{NS}(\omega,Q_t^{\prime 2})+ 
\int_{Q_t^2}^{\infty} {dQ_t^{\prime 2}\over 
Q_t^{\prime 2}}\left({Q_t^2 \over Q_t^{\prime 2}}\right)^{\omega}
\Delta \bar f_q^{NS}(\omega,Q_t^{\prime 2})\right]
\label{dleqm}
\end{equation}
For fixed coupling $\tilde \alpha_s$  equation (\ref{dleqm})
 can be solved analytically.  
Assuming for simplicity that the inhomogeneous term is independent 
of $Q_t^2$ (i.e. that $\Delta \bar f_{q0}^{NS}(\omega,Q_t^2) = C(\omega)$ )
we get the following solution of  eq.(\ref{dleqm}): 
\begin{equation}
\Delta \bar f_q^{NS}(\omega,Q_t^2)=C(\omega)R(\tilde \alpha_s,  \omega) 
\left({Q_t^2\over Q_0^2}\right)^{ \gamma^{-}(\tilde \alpha_s,  \omega)}
\label{solm}
\end{equation}
where 
\begin{equation}
\gamma^{-}(\tilde \alpha_s, \omega) = {\omega - \sqrt{\omega^2 - 4 \tilde 
 \alpha_s}\over 2}
\label{anomd}
\end{equation}
and
\begin{equation}
R(\tilde \alpha_s,  \omega)= {\omega \gamma^{-}(\tilde \alpha_s, \omega)\over 
\tilde \alpha_s}. 
\label{r}
\end{equation}
Equation (\ref{anomd}) defines the anomalous dimension of the 
 moment of the non-singlet quark distribution in which 
 the double logarithmic $ln(1/x)$ terms i.e. the powers of ${\alpha_s \over 
\omega^2}$ have been resummed to all orders.  It can be seen from (\ref{anomd}) 
that this anomalous dimension has a (square root) branch point singularity 
at $\omega=
\bar \omega$ where 
\begin{equation}
\bar \omega= 2 \sqrt{\tilde \alpha_s}. 
\label{barom}
\end{equation} 
This singularity will of course be also present in the moment function $
\Delta \bar f_q^{NS}(\omega,Q_t^2)$ itself. It should be noted that in contrast to the 
BFKL singularity whose position above unity was proportional to $\alpha_s$,  
$\bar \omega$ is proportional to $\sqrt{\alpha_s}$ - this being the 
straightforward consequence of the fact that  equation (\ref{dleqm}) 
sums double logarithmic terms $({\alpha_s\over \omega^2})^n$. 
This singularity gives the following contribution to the 
non-singlet quark distribution $\Delta f_q^{NS}(x,Q_t^2)$ at small 
$x$:  
\begin{equation} 
\Delta f_q^{NS}(x,Q_t^2) \sim {x^{-\bar \omega}\over ln^{3/2}(1/x)}. 
\label{smxns}
\end{equation}

The introduction of the running coupling effects in  equation (\ref{dleqm})
turns the branch point singularity into the series of poles which accumulate 
at $\omega=0$ \cite{JK2}.  The numerical analysis of the corresponding 
integral equation,  
with the running coupling effects taken into account,  
gives an effective slope ,  
\begin{equation}
\lambda(x,Q_t^2)={dln \Delta f_q^{NS}(x,Q_t^2)\over d ln(1/x)}
\label{slope}
\end{equation}
with magnitude $\lambda(x,Q_t^2) \approx 0.2 - 0.3$ at small $x$ 
\cite{JK3}.   
The result of this estimate suggests that a reasonable extrapolation 
of the (non-singlet) polarized quark densities would be to assume an  
$x^{-\lambda}$ behaviour with $\lambda \approx 0.2 - 0.3$.   Similar 
 extrapolations of the spin-dependent quark 
distributions towards the small $x$ region have  
 been assumed in  several recent parametrizations of parton densities 
\cite{BS,GRVOG,GS,BV}.  
The perturbative QCD effects become significantly amplified for the 
singlet spin structure function due to  mixing with the gluons.  
The simple ladder equation may not however be  applicable 
for an accurate description of the double logarithmic terms in 
the polarized gluon distribution $\Delta G$ \cite{BERSING}.
The small $x$ behaviour of the spin dependent 
structure function $g_1$ has also been disscussed in refs. \cite{BASLO,RGRF}.
\\
\section*{3. Dedicated measurements probing the QCD pomeron in deep inelastic 
scattering at low $x$}

It is expected that absence of transverse momentum ordering along the gluon 
chain which leads to the correlation between the increase of the 
structure function  with  decreasing $x$ and the diffusion 
of transverse momentum should reflect itself in the behaviour of 
less inclusive 
quantities than the structure function $F_2(x,Q^2)$.  The dedicated 
measurements of the low $x$ physics which are particularly sensitive 
to this correlation are the deep inelastic scattering plus energetic forward 
jet events
\cite{MJET,BJET,WKJET,KMSJET,BJET1,EWELINA}, transverse energy flow 
in deep inelastic scattering \cite{KMSET,KUHLEN},  
production of jets separated by the large 
rapidity gap \cite{DDUCA,JAMESJ} and dijet 
production in photoproduction \cite{DICKJET} and in deep inelastic scattering 
\cite{AGKM}. 
Complementary measurement to deep inelastic scattering plus forward jet is the 
deep inelastic scattering accompanied by the forward prompt photon or 
forward prompt $\pi^0$ \cite{KLM1,KLM2}.\\

In principle  deep inelastic lepton scattering containing a measured   
energetic jet 
can provide a very clear test of the BFKL dynamics at low $x$
\cite{MJET,BJET,WKJET,KMSJET,BJET1,EWELINA}.  
The idea is to study deep inelastic ($x,Q^2$) events which 
contain an identified jet ($x_j,k_{Tj}^2$) where 
$x<<x_j$ and $Q^2 \approx k_{Tj}^2$.  Since we choose events with 
$Q^2 \approx k_{Tj}^2$ the leading order QCD evolution  (from $k_{Tj}^2$ to 
$Q^2$) is neutralized and attention is focussed on the small $x$, or rather 
small $x/x_j$ behaviour.  The small $x/x_j$ behaviour
of jet production is generated  by the gluon radiation. Choosing the 
configuration $Q^2 \approx k_{Tj}^2$ we eliminate by definition  
gluon emission which corresponds to strongly ordered transverse momenta 
i.e.  that emission which is responsible for the LO QCD evolution.  
The measurement of jet production in this configuration may therefore test 
more directly the $(x/x_j)^{-\lambda}$ behaviour which is generated 
by the BFKL equation where the transverse momenta are not ordered. \\

  The 
differential cross section for the deep inelastic + 
jet process depicted in Fig. 3a is given by \cite{KMS2}
\begin{equation}
\frac{\partial \sigma_{j}}{\partial x \partial Q^{2}} = \int dx_{j}
\int dk_{jT}^{2} \frac{4 \pi \alpha ^{2}}{xQ^{4}} \left[
\left(1-y \right) \frac{\partial F_{2}}{\partial x_{j} \partial
k_{jT}^{2}} + \frac{1}{2} y^{2} \frac{\partial \left(2xF_{T} 
\right)}{\partial x_{j} \partial k_{jT}^{2}} \right] 
\label{eq:a2}
\end{equation}
where the differential structure functions have the following
 form
\begin{equation}
\frac{\partial^2 F_{i}}{\partial x_{j} \partial k_{jT}^{2}} =
\frac{3 \alpha_S \left(k_{jT}^{2} \right) }{\pi k_{jT}^{4}}
\sum_{a} f_a \left(x_{j},k_{jT}^{2} \right) \Phi_i
\left(\frac{x}{x_{j}},k_{jT}^{2},Q^{2} \right)
\label{eq:a3}
\end{equation}
for $i=T,L$. Assuming $t$-channel pole dominance the sum over 
the parton distributions is given by
\begin{equation}
\sum_{a} f_{a} = g+\frac{4}{9} \left(q+ \bar q \right).
\label{eq:a4}
\end{equation}
Recall that these parton distributions
are to be evaluated at $(x_{j},k_{jT}^2)$ where 
they are well-known from the global analyses, so there are no ambiguities
arising from a non-perturbative input.\\

The functions $\Phi_i (x/x_j, k_{jT}^{2}, Q^2)$
in (\ref{eq:a3}) describe the virtual $\gamma$ + virtual gluon fusion process 
including the ladder formed from the gluon chain of Fig. 3a.  They can be 
obtained by solving the BFKL equation

$$
\Phi_i (z, k_T^2, Q^2)  =  \Phi_i^{(0)} (z, k_T^2, Q^2)+
$$
\begin{equation}
\bar \alpha_s \int_z^1{dz^{\prime}\over 
z^{\prime}} \int {d^2 q\over \pi q^2}
\left[{k_T^2 \over (\mbox{\boldmath $q$}+
\mbox{\boldmath $k_T$})^2} 
\Phi_i(z^{\prime},(\mbox{\boldmath $q$}+
\mbox{\boldmath $k_T$})^2,Q^2))-\Phi_i(z^{\prime},k_T^2,Q^2)
\Theta(k_T^2-q^2)\right]
\label{eq:a6}
\end{equation}

The inhomogeneous or driving terms $\Phi_{i}^{(0)}$ correspond
to the sum of the quark box and crossed-box contributions.
For small $z$ we have
For small $z$ we have
\begin{equation}
\Phi_i^{(0)} (z, k_T^2, Q^2) \; \approx \; \Phi_i^{(0)} (z = 0,
k_T^2, Q^2) \; \equiv \; \Phi_i^{(0)} (k_T^2, Q^2).
\label{eq:aa66}
\end{equation}

We evaluate the $\Phi_i^{(0)}$ by expanding the four momentum in terms of the 
basic light-like four momenta $p$ and $q^\prime \equiv q + xp$.  For example, 
the quark momentum ${\kappa}$ in the box (see Fig. 3a) has the Sudakov 
decomposition
$$
\kappa \; = \; \alpha p \: - \: \beta q^\prime \: + \: \mbox{\boldmath $\kappa$}_T.
$$
We carry out the integration over the box diagrams, subject to the quark 
mass-shell constraints, and find
\begin{eqnarray}
\Phi_T^{(0)} (k_T^2, Q^2) & = & 2 \: \sum_q \: e_q^2 \:
\frac{Q^2}{4 \pi^2} \: \alpha_S \: \int_0^1 \: d \beta \: \int \:
d^2 \kappa_T \left [\beta^2 \: + \: (1 - \beta)^2 \right ] \left (
\frac{\kappa_T^2}{D_1^2} \: - \: \frac{\mbox{\boldmath $\kappa$}_T .
(\mbox{\boldmath $\kappa$}_T - \mbox{\boldmath $k$}_T)}{D_1 D_2}
\right ) \nonumber \\
& & \nonumber \\
\Phi_L^{(0)} (k_T^2, Q^2) & = & 2 \sum_q \: e_q^2 \;
\frac{Q^4}{\pi^2} \: \alpha_S \: \int_0^1 \: d \beta \: \int \:
d^2 \kappa_T \: \beta^2 (1 - \beta)^2 \; \left ( \frac{1}{D_1^2} \:
- \: \frac{1}{D_1 D_2} \right ).
\label{eq:a7}
\end{eqnarray}
where the denominators $D_{i}$ are of the form
\begin{eqnarray}
D_1 & = & \kappa_T^2 \: + \: \beta (1 - \beta) \: Q^2 \nonumber \\
& & \\
D_2 & = & (\mbox{\boldmath $\kappa$}_T - \mbox{\boldmath $k$}_T)^2 \:
+ \: \beta (1 - \beta) \: Q^2, \nonumber
\label{eq:a8}
\end{eqnarray}
assuming massless quarks.\\

If the QCD coupling $\alpha_S$ is fixed we can solve the BFKL
equation (\ref{eq:a6}) and obtain an analytic expression for the leading
small $z$ behaviour of the solution, including its normalisation. 
Omitting the Gaussian diffusion
factor in $\ln \left(k_T^{2}/Q^{2} \right)$ we find
\begin{eqnarray}
\Phi_T (z, k_T^2, Q^2) & = & \frac{9 \pi^2}{512} \: \frac{2 \sum
e_q^2 \alpha_S^{\frac{1}{2}}}{\sqrt{21 \zeta (3)/2}} \; (k_T^2
Q^2)^{\frac{1}{2}} \; \frac{z^{-\lambda_{BFKL}}}{\sqrt{\ln (1/z)}}
\; \left [ 1 \: + \: {\cal O} \: \left ( \frac{1}{\ln (1/z)}
\right ) \right ] \nonumber \\
& & \\
\Phi_L (z, k_T^2, Q^2) & = & \frac{2}{9} \: \Phi_T (z, k_T^2, Q^2)
\nonumber
\label{eq:a9}
\end{eqnarray}

Singular small $z$ behaviour of the functions $\Phi_i$ leads to 
increase of the cross section for forward jet production with decreasing $x$. 
Recent H1   results concerning  deep inelastic plus jest events 
are consistent with this prediction {\cite{BJET1,EWELINA}. \\

As it has been mentioned above a complementary measurement  
to deep inelastic scattering plus forward jet is the 
deep inelastic scattering accompanied by the forward prompt photon or 
forward prompt $\pi^0$ \cite{KLM1,KLM2}.\\ 

The prompt photon production in deep-inelastic
scattering at small $x$ is the process
\begin{equation}
\lq\lq \gamma" \: + \: p \; \rightarrow \; \gamma (x_\gamma,
k_{\gamma T}) \: + \: X,
\label{eq:aa3}
\end{equation}
sketched in Fig. 3b , in which the photon is identified in the
final state.  Deep inelastic events with small $x$ and large
$x_\gamma$ offer an opportunity to identify the effects of the
BFKL resummation of the $\alpha_S \ln (x_\gamma/x)$
contributions, which arise from the sum over the real and virtual
gluon emissions, such as the one depicted in Fig. 3b.  In
analogy to the DIS + jet process, the advantage of process
(\ref{eq:aa3}) is that the outgoing photon acts as a trigger to
select events in which the deep-inelastic scattering occurs off a
quark in a kinematic region where its distribution, $q (x_q,
k_{\gamma T}^2)$, is known.

The differential structure functions for this process may be
written in the form
\begin{equation}
\frac{\partial^2 F_i}{\partial x_\gamma \partial k_{\gamma T}^2}
\; = \; \int \: \frac{d^2 k_{gT}}{\pi k_{gT}^4} \: \int \:
\frac{dx_q}{x_q} \: \Phi_i \left (\frac{x}{x_q}, k_{gT}^2, Q^2
\right ) \sum_q e_q^2 \: \left [ q (x_q, k_{\gamma T}^2) \: + \:
\overline{q} (x_q, k_{\gamma T}^2) \right ] \; \frac{|{\cal
M}|^2}{z_\gamma z_q^\prime}
\label{eq:aa4}
\end{equation}
for $i = T, L$.  The variables are indicated on Fig. 3b and
are defined below.  The subprocess $q (k_q) + g (k_g)
\rightarrow
q (k_q^\prime) + \gamma (k_\gamma)$ (and also $\overline{q}g
\rightarrow \overline{q}\gamma$) is described by the two Feynman
diagrams shown in Fig. 4.  It has matrix element squared
\begin{equation}
|{\cal M}|^2 \; = \; 2 \: C_2 (F) \; \frac{\alpha_S (k_{\gamma
T}^2)}{2
\pi} \; \frac{\alpha}{2 \pi} \; \frac{k_{gT}^2 [1 \: + \: (1 -
z_\gamma)^2]}{\hat{s} (- \hat{u})}
\label{eq:a5}
\end{equation}
where $C_2 (F) = 4/3$ and the invariants $\hat{s}$ and $\hat{u}$
are
\begin{equation}
\hat{s} \; = \; (k_\gamma \: + \: k_q^\prime)^2, \;\;\;\;\;
\hat{u} \; = \; (k_q \: - \: k_\gamma)^2.
\label{eq:aa6}
\end{equation}
The fractional momenta
$x_i$ in (\ref{eq:aa4}) are defined by the Sudakov decomposition
of the particle 4-momenta
\begin{equation}
k_i \; = \; x_i  p^\prime \: + \: \beta_i q^\prime \: + \:
\mbox{\boldmath $k$}_{iT}
\label{eq:aa7}
\end{equation}
  The variables $z_i$ in
(\ref{eq:aa4}) and (\ref{eq:a5}) are given by
\begin{equation}
z_i \; = \; x_i/x_q.
\label{eq:aa9}
\end{equation}
The 4-momenta of the outgoing photon and quark jet satisfy the
on-mass-shell condition $k_i^2 = 0$ which gives
\begin{equation}
\beta_i \; = \; \frac{x}{x_i} \: \frac{k_{iT}^2}{Q^2}
\label{eq:b9}
\end{equation}
for $i = \gamma$ or the outgoing quark $q^\prime$.\\

In Fig. 5 we show the integrated cross section for prompt
forward photon
production as a function of $x$ for three different $Q^2$ bins: 
20-30, 30-40 and 40-50 GeV$^2$ respectively.  The cross-section 
was calculated imposing various acceptance cuts.    We compare the
predictions for the case where the BFKL small $x$ resummation is
incorporated with those where the gluon radiation is neglected. 
The $x$ dependence of the cross section is driven by the small
$z$ behaviour of the $\Phi_i$.  The results show a strong
enhanced increase with decreasing $x$ which is characteristic of
the effect of soft gluon resummation. \\

In order to
calculate the cross section for DIS + $\pi^0$ production (see Fig. 6) we have
to convolute the DIS + jet cross section with the $\pi^0$
fragmentation functions. We obtain
\begin{eqnarray}
\frac{\partial \sigma_{\pi}}{\partial x_{\pi} \partial k_{\pi T}} = 
\int_{x_{\pi}}^{1} dz \int dx_{j} \int dk_{j T}^{2}
\left[ \frac{\partial \sigma_{g}}{\partial x_{j} \partial k_{jT}^{2}} 
D_{g}^{\pi^0} \left( z,k_{\pi T}^{2} \right) + \sum_{q}
\left( \frac{\partial \sigma_{q}}{\partial x_{j} \partial k_{jT}^{2}}
D_{q}^{\pi^0} \left( z,k_{\pi T}^{2} \right) \right) \right] 
\times \nonumber \\
\times \delta \left(x_{\pi} - zx_{j} \right)
\delta \left(k_{\pi T} - zk_{jT} \right) 
\label{eq:ab14}
\end{eqnarray}
where the sum over $q$ runs over all quark and antiquark flavours.
The partonic differential cross sections can be obtained from 
(\ref{eq:a2}) and (\ref{eq:a3})
by substituting for the sum over the parton distributions
$\sum_{a} f_{a}$ either the gluon distribution $g$ or the quark
or antiquark distribution $\frac{4}{9} q$ or $\frac{4}{9} \bar q$
respectively. \\

The results for the integrated cross-section for forward $\pi^0$ 
production in deep inelastic scattering are summarized in Fig. 7. \\

Conceptually similar process to deep inelastic  
scattering plus forward jet is that of the two-jet production 
separated by a large rapidity gap $\Delta y$ in hadronic 
collisions or in photoproduction
\cite{DDUCA,JAMESJ}.  
Besides the characteristic $exp(\lambda \Delta y)$ dependence 
of the two-jet cross-section one expects significant 
weakening of the azimuthal back-to-back correlations 
of the two jets.  This is the direct consequence of the 
absence of transverse momentum ordering along the gluon  
chain.     
Another measurement which should be sensitive to the QCD 
dynamics at small $x$ is that of the transverse energy flow in deep inelastic 
lepton scattering in the central region away from the 
current jet and from the proton remnant  
\cite{KMSET}.  
The  BFKL dynamics predicts 
in this case a substantial amount of transverse energy  which should 
increase with decreasing $x$.  The experimental data 
are consistent with this theoretical expectation \cite{KUHLEN}.  
Absence of transverse momentum ordering  also implies weakening of the 
back-to-back azimuthal correlation of dijets produced close 
to the photon fragmentation region \cite{DICKJET,AGKM}. \\

Interesting insight into the BFKL equation (\ref{bfkl}) can also be obtained by 
studying the exclusive intermediate states of the gluon ladder which 
contains the multijet structure \cite{KLEWM}. 
Jet structure is embodied in the BFKL equation via real gluon
emission from the gluon chain prior to its interaction with the
photon probe (which takes place through the usual fusion
subprocess
$\gamma g \rightarrow q\overline{q}$).  An observed jet is
defined by a resolution parameter $\mu$ which specifies the
minimum transverse momentum that must be carried by the emitted
gluon for it to be detected.  For realistic observed jets in the 
experiments at HERA, the lowest reasonable choice for the
resolution cut-off parameter $\mu$ appears to be about $\mu =
3.5$ GeV. If an emitted gluon has transverse momentum $q_T < \mu$ then the
radiation is said to be unresolved.  The unresolved radiation
must be treated at the same level as the virtual corrections to
ensure that the singularities as $q_T^2 \rightarrow 0$ cancel in
the $q_T^2$ integration.  To do this we first rewrite the BFKL
equation (\ref{bfkl}) in the symbolic form
\begin{equation}
f \; = \; f^{(0)} \: + \: \int_0^y \: dy^\prime \: K \otimes f
(y^\prime),
\label{eq:ac66}
\end{equation}
where $\otimes$ denotes the convolution over $q$ and $y=ln(1/x)$.  We divide
the real gluon emission contribution into resolved
and unresolved parts using the identity
\begin{equation}
\Theta (q^2 - \mu^2) \: + \: \Theta (\mu^2 - q^2) \; = \; 1,
\label{eq:ac7}
\end{equation}
where the first term denotes the real resolved emission and the 
second the real unresolved emission. We then combine the unresolved 
component with the virtual
contribution.  That is
\begin{equation}
f \; = \; f^{(0)} \: + \: \int_0^y \: dy^\prime \: (K_R + K_{UV})
\: \otimes \: f (y^\prime),
\label{eq:ac8}
\end{equation}
where the kernel $K_R$ for the {\it resolved} emissions with $q
> \mu$ is given by
\begin{equation}
K_R \: \otimes \: f(y^\prime) \; = \; \overline{\alpha}_S (Q_t^2)
\: Q_t^2 \: \int \: \frac{d^2 q}{\pi q^2} \: \Theta (q^2 -
\mu^2) \: \frac{1}{Q_t^{\prime 2}} \: f (y^\prime, Q_t^{\prime
2}),
\label{eq:ac9}
\end{equation}
while $K_{UV}$, the combined {\it unresolved} and {\it virtual}
part of the kernel, satisfies
\begin{equation}
K_{UV} \: \otimes \: f (y^\prime) \; = \; \overline{\alpha}_S
(Q_t^2) \: \int \: \frac{d^2 q}{\pi q^2} \: \left [
\frac{Q_t^2}{Q_t^{\prime 2}} \: f (y^\prime, Q_t^{\prime 2}) \:
\Theta (\mu^2 - q^2) \: - \: f (y^\prime, Q_t^2) \: \Theta
(Q_t^2 - q^2) \right ],
\label{eq:a1c0}
\end{equation}
with $Q_t^{\prime 2} \equiv |\mbox{\boldmath $q$} +
\mbox{\boldmath $Q$}_t |^2$.  After resumming the unresolvable real emission 
and virtual correction terms the BFKL equation can be rearranged into the 
following form: 
\begin{equation}
f (y) \; = \; \hat{f}^{(0)} (y) \: + \: \int_0^y \: dy^\prime \:
\hat{K} \: \otimes \: f (y^\prime)
\label{eq:ac15}
\end{equation}
where the driving term has become
\begin{equation}
\hat{f}^{(0)} (y) \; = \; \int_0^y \: dy^\prime \: e^{(y -
y^\prime) K_{UV}} \: \otimes \: \frac{\partial f^{(0)}}{\partial
y^\prime}
\label{eq:ac16}
\end{equation}
and the new kernel
\begin{equation}
\hat{K} \; = \; e^{(y - y^\prime) K_{UV}} \: \otimes \: K_R.
\label{eq:ac17}
\end{equation}
Iteration of the equation (\ref{eq:ac15}) generates decomposition of the unintegrated distribution 
$f$ into the sum of contributions with different numbers of $resolved$ gluon 
jets.  That is 
\begin{equation}
f(y)= \hat f^{(0)}(y) + \sum_{n=1}^{\infty} f^n(y)
\label{decomp}
\end{equation} 
where 
$$f^1(y)=\int_0^ydy^{\prime}\hat K \otimes \hat f^{(0)}(y^{\prime}) $$
and 
\begin{equation} 
f^n (y) \; = \; \int_0^y \: dy^\prime \: \hat{K} \: \otimes \:
f^{n - 1} (y^\prime)
\label{eq:ac19}
\end{equation}
for $n>1$.  Using the $k_t$ factorisation theorem (\ref{ktfac}) we get 
the contributions of different numbers of resolved jets to the structure 
functions which are illustrated in Fig.8.  
  
\section*{4. Deep inelastic diffraction}
Important process which is sensitive to the small $x$ dynamics 
is  the deep inelastic diffraction \cite{ZDIF,H1DIF}. 
Deep inelastic diffraction in $ep$ inelastic scattering is a process: 
\begin{equation}
e(p_e)+p(p) \rightarrow e^{\prime}(p_e^{\prime}) + X +p^{\prime}(p^{\prime})
\label{disdif}
\end{equation}
where there is a large rapidity gap between the recoil proton 
(or excited proton) and the hadronic system $X$.  
To be precise   
process (\ref{disdif}) reflects the diffractive disssociation 
of the virtual photon.  Diffractive dissociation is described by the following 
kinematical variables: 
\begin{equation}
\beta={Q^2\over 2 (p-p^{\prime})q}
\end{equation}
\begin{equation}
x_P={x\over \beta}
\end{equation}
 \begin{equation}
t= (p-p^{\prime})^2. 
\label{difv}
\end{equation}
Assuming that  diffraction dissociation is dominated by the pomeron 
exchange  and that the pomeron is described by 
a Regge pole one gets the following factorizable expression for the 
diffractive structure function \cite{DOLADIF,ORSAYDIF,CTEQD,JAMESD,KGBK}:
\begin{equation}
{\partial F_2^{diff}\over \partial x_P \partial t}= f(x_P,t)F_2^P(\beta,Q^2,t)
\label{difsf}
\end{equation}
where the "flux factor" $f(x_P,t)$ is given by the following formula
:
\begin{equation}
f(x_P,t)=N{B^2(t)\over 16\pi} x_P^{1-2\alpha_P(t)}
\label{flux}
\end{equation}
with $B(t)$ describing the pomeron coupling to a proton and $N$ being the 
normalisation factor.  The function $F_2^P(\beta,Q^2,t)$
is the pomeron structure function which in the (QCD improved) parton model 
is related in a standard way to the quark and antiquark distribution 
functions in a pomeron.  
\begin{equation}
F_2^P(\beta,Q^2,t)=\beta \sum e_i^2[q_i^P(\beta,Q^2,t)+ \bar 
q_i^P(\beta,Q^2,t)]
\label{f2pom}
\end{equation} 
with $q_i^P(\beta,Q^2,t)=\bar q_i^P(\beta,Q^2,t)$.  The variable 
$\beta$ which is the Bjorken scaling variable appropriate for 
deep inelastic lepton-pomeron "scattering",  has the meaning of the 
momentum fraction of the pomeron carried by the 
probed quark (antiquark).  The quark distributions in a pomeron 
are assummed to obey the standard Altarelli-Parisi evolution equations: 
\begin{equation}
Q^2{\partial q^P\over \partial Q^2}=P_{qq} \otimes q^P + P_{qg} \otimes g^P
\label{app}
\end{equation}
with a similar equation for the evolution of the gluon distribution 
in a pomeron.  The first 
term on the right hand side of the eq. (\ref{app}) becomes negative 
at large $\beta$,  while the second term remains positive and 
is usually very small at large $\beta$ unless the gluon distributions 
are large and have a hard spectrum.\\

 The data suggest that the slope 
of $F_2^P$ as the function of $Q^2$ does not change sign even 
at relatively large values of $\beta$.  This favours the hard 
gluon spectrum in a pomeron \cite{CAPHG,KGBJP}, and   
should be contrasted with the behaviour of the structure function 
of the proton which, at large $x$, decreases with increasing $Q^2$.   
The data on inclusive diffractive production  favour the soft pomeron 
with relatively low intercept.  
 The diffractive production of vector mesons 
 seems to require a "hard" pomeron contribution 
\cite{HALINA} . It has also 
been pointed out  that the factorization property 
(\ref{difsf}) may 
not hold in models based 
entirely on perturbative QCD when the pomeron is represented 
by the BFKL ladder \cite{KOLYA,BARTELSD}.  The factorization does not also 
hold when the exchange of the secondary Regge poles besides pomeron 
becomes important \cite{KGBSL,PWAW}.  The contribution of secondary Reggeons 
is expected to be significant at moderately small $x_P$ and small values of 
$\beta$.  The recent HERA results show violation of factorization in this 
region \cite{PWAW}. Finally let us point out that 
there exist also models of deep inelastic diffraction which do not 
rely on the pomeron exchange picture \cite{BUCHMD,EIR}.\\

\section*{4. Summary and conclusions}  
   
Perturbative QCD predicts  indefinite increase of gluon 
distributions 
with decreasing $x$ which generates similar increase of the structure 
functions through the 
$g \rightarrow q \bar q$ transitions. 
The indefinite growth of parton distributions cannot go on forever 
and has to be eventually stopped by parton screening which leads 
to the parton saturation.  Most probably however 
the saturation limit is still irrelevant for the small $x$ region 
which is now being probed at HERA. Besides discussing the theoretical and 
phenomenological issues related to the description of the structure 
function $F_2$ at low $x$ we have also emphasised 
the role of studying the hadronic final state in deep inelastic scattering 
for probing the QCD pomeron. Finally let us point out that the  recent experiments at HERA  cover very 
broad range of $Q^2$ including the region of low and moderately 
large values of $Q^2$.  Analysis of the structure functions in
this  transition region is very interesting \cite{BBJK}               
 and may help to understand 
possible relation (if any) between the soft and hard pomerons.

\medskip\medskip\medskip                
     
\section*{Acknowledgments}
I thank the organizers of the School for organizing an excellent meeting. 
 I thank Barbara Bade\l{}ek, Krzysztof Golec-Biernat, Sabine Lang, Claire Lewis, 
 Alan Martin and Peter Sutton 
for the very enjoyable research collaboration on problems presented 
in these lectures. 
This research has been supported in part by 
 the Polish State Committee for Scientific Research grant 2 P03B 231 08  and 
the EU under contracts n0. CHRX-CT92-0004/CT93-357.\\

\section*{Figure Captions}
\begin{description}
\item{1.} The exponent $\lambda_{BFKL}$ of the $x^{- \lambda_{BFKL}}$
behaviour of the gluon distribution obtained by solving the BFKL
equation (a) with (continuous curve) and (b) without (dashed
curve) the  constraint $q^2 < Q_t^2x^{\prime}/x$ imposed, 
as a function of (fixed)
$\overline{\alpha}_S \equiv 3 \alpha_S/\pi$.  The dashed curve is
$\lambda_{BFKL} = \overline{\alpha}_S 4 \ln 2$.  The dotted curve (c) is
the value of the exponent that is obtained if we keep only the
next-to-leading order modification due to the 
constraint $q^2 < Q_t^2x^{\prime}/x$.  (From \cite{KMSGLU}). 

\item{2.}The continuous and dashed curves correspond to the values of the proton
structure function $F_2$ obtained from the R$_1$ and R$_2$
sets of partons (which have, respectively, QCD couplings corresponding
to $\alpha_s(M_Z^2) = 0.113$ and $0.120$) at twelve values
of $x$ chosen to be the most appropriate for the new HERA data.
For display purposes we add $0.5(12-i)$ to $F_2$ each time
the value of $x$ is decreased, where $i=1,12$.
For comparison the dotted curves show the prediction obtained from the 
GRV set of partons \cite{GRV}.
The experimental data are assigned to the $x$ value which is
closest to the experimental $x$ bin. (From \cite{MRSR}). 

\item{3.} Diagrammatic representation of (a) a
deep-inelastic + forward jet event, and (b) a deep-inelastic $(x,
Q^2)$ + forward identified photon $(x_\gamma, k_{\gamma T})$
event.(From \cite{KLM1}). 

\item{4.} The Feynman diagrams describing the $gq
\rightarrow \gamma q$ subprocess embodied in the DIS + $\gamma$
diagram shown in Fig. 3b. (From \cite{KLM1}). 

\item{5.} The cross section, $\langle \sigma \rangle$ in
pb, for deep inelastic + photon events integrated over $\Delta
x = 2 \times 10^{-4}$, $\Delta Q^2 = 10$ GeV$^2$ bins which are
accessible at HERA, and subject to various acceptance cuts. 
The $x$ dependence is shown for three different
$\Delta Q^2$ bins, namely (20,30), (30,40) and (40,50) GeV$^2$. 
The $\langle \sigma \rangle$ values are plotted at the central
$x$ value in each $\Delta x$ bin and joined by straight lines. 
The continuous curves show $\langle \sigma \rangle$ calculated
with $\Phi_i$ determined from the BFKL equation, whereas the 
dashed
curves are obtained just from the driving terms $\Phi_i^{(0)}$,
i.e.\ from the quark box.  For clarity a vertical line links the
pair of curves belonging to the same $\Delta Q^2$ bin. (From \cite{KLM1}). 

\item{6.} Diagrammatic representation of (a) a
deep inelastic + forward jet event, and (b) a deep inelastic $(x,
Q^2)$ + identified forward $\pi^{0}$ $(x_\pi, k_{\pi T})$
event.(From \cite{KLM2}). 

\item{7.} The cross section, $\langle \sigma \rangle$ in pb,
for deep inelastic + $\pi^{0}$ events integrated over bins of
size $\Delta x = 2 \times 10^{-4}$, $\Delta Q^2 = 10$ GeV$^2$
which are accessible at HERA for $\pi^{0}$'s with transverse
momentum $3 < k_{\pi T} < 10$ GeV and subject to various acceptance cuts. 
The $\langle \sigma \rangle$ values are plotted at 
the central $x$ value in each $\Delta x$ bin and joined by 
straight lines. The $x$ dependence is plotted for three different
$\Delta Q^2$ bins, namely (20,30), (30,40) and (40,50) GeV$^2$.
The continuous curves show $\langle \sigma \rangle$ calculated
with $\Phi_i$ obtained from the BFKL equation. The corresponding
$\langle \sigma \rangle$ values calculated neglecting soft gluon 
resummation and just using the quark box approximation
$\Phi_{i} = \Phi_{i}^{(0)}$ are plotted as dashed curves.
For clarity a dotted vertical lines joins each pair of curves
belonging to the same $\Delta Q^2$ bin. (From \cite{KLM2}). 

\item{8.} The decomposition of the proton structure function $F_2 (x, Q^2)$ 
into contributions coming from different numbers of resolved gluon jets for 
experimentally accessible values of the resolution parameter $\mu = 3.5$ and 
6 GeV.  The decomposition is shown as a function of $x$ for $Q^2 = 10$ GeV$^2$. 
(From \cite{KLEWM}). 
\end{description}
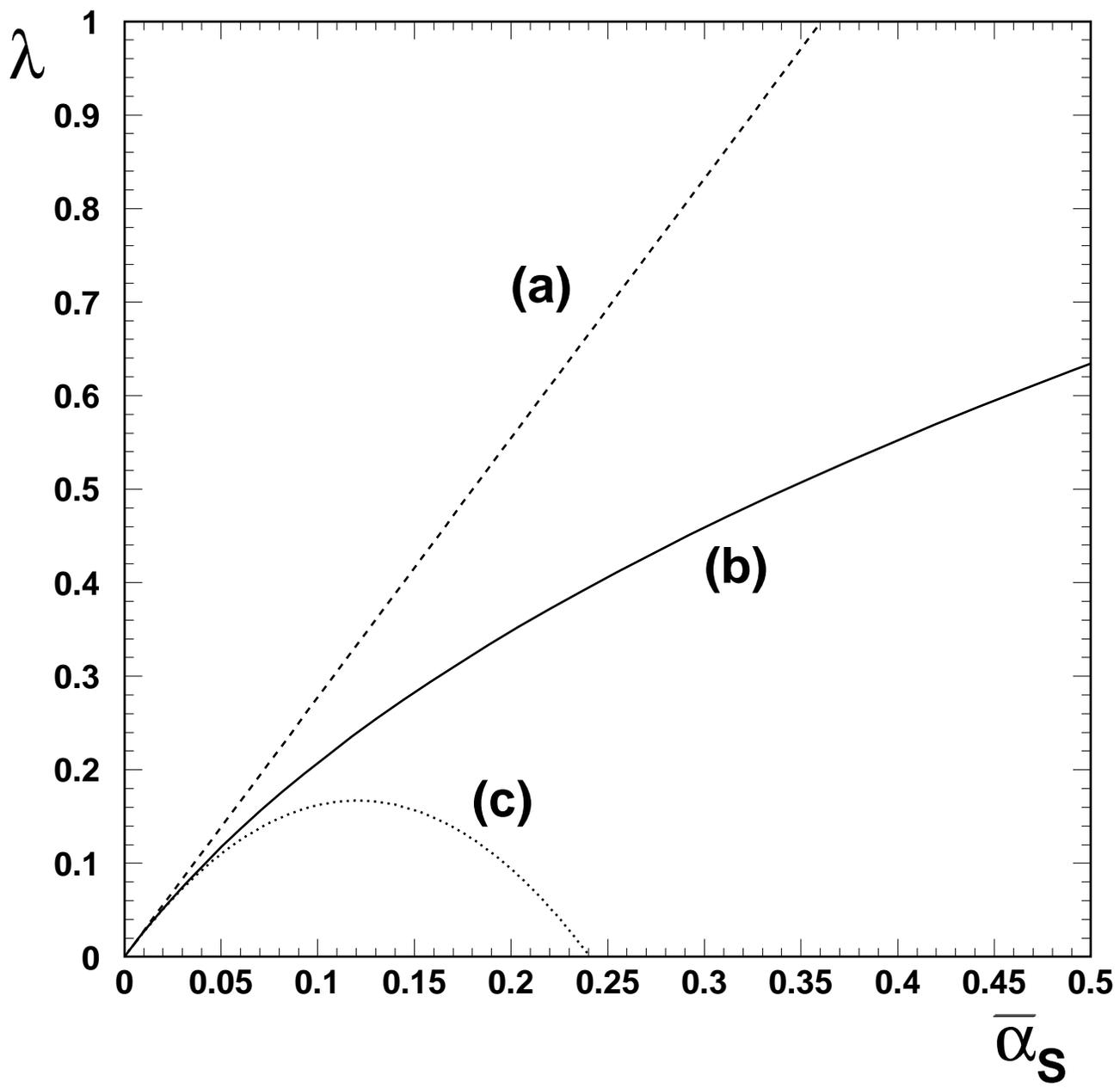
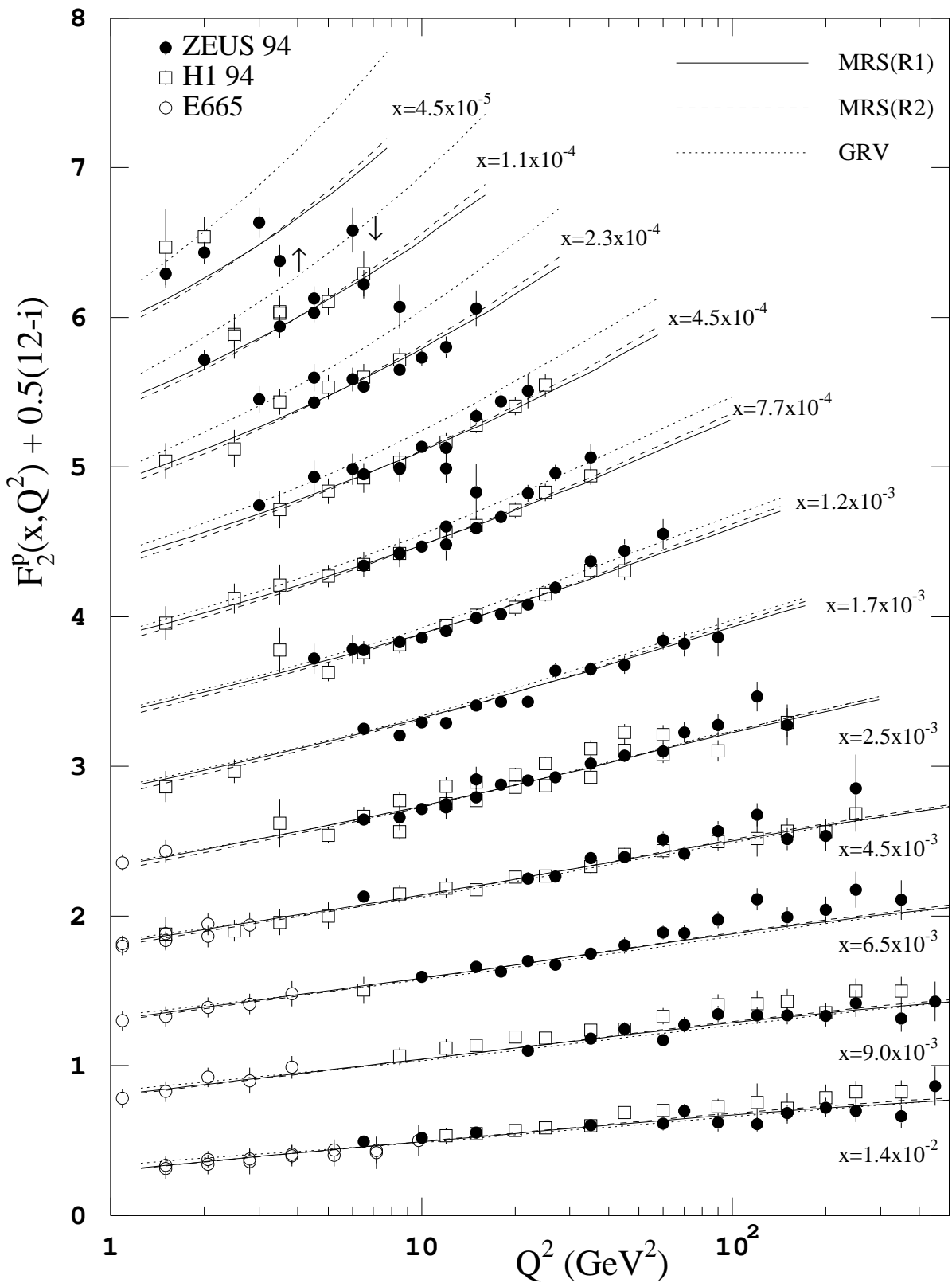
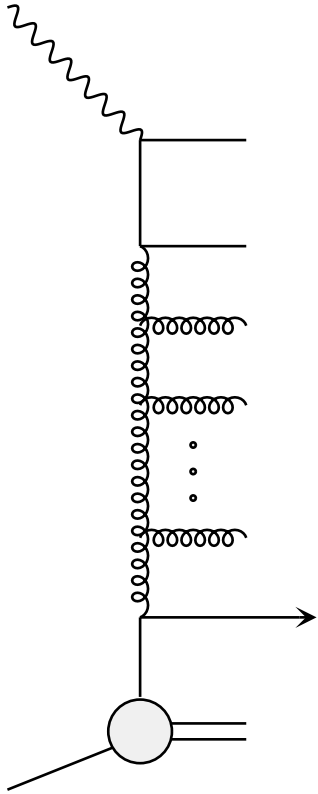
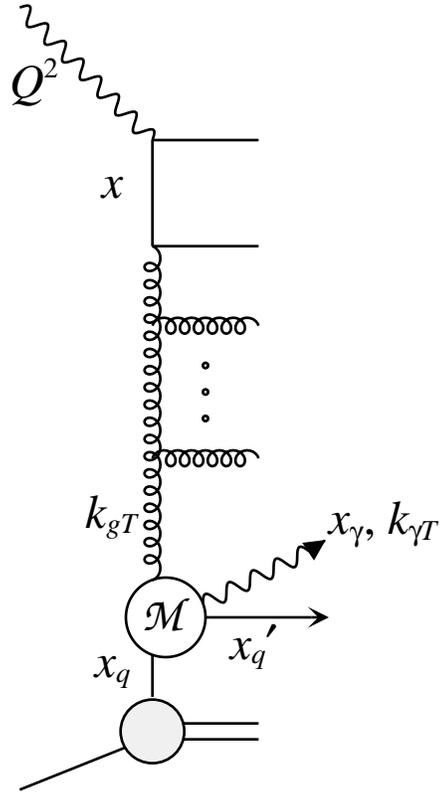
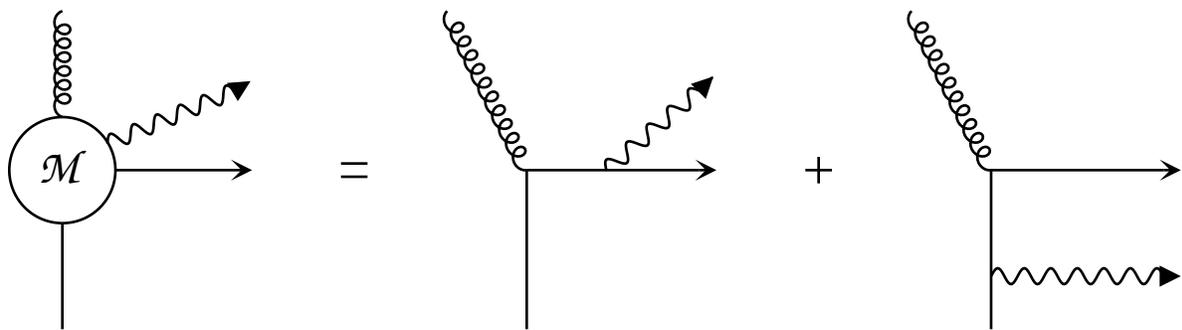
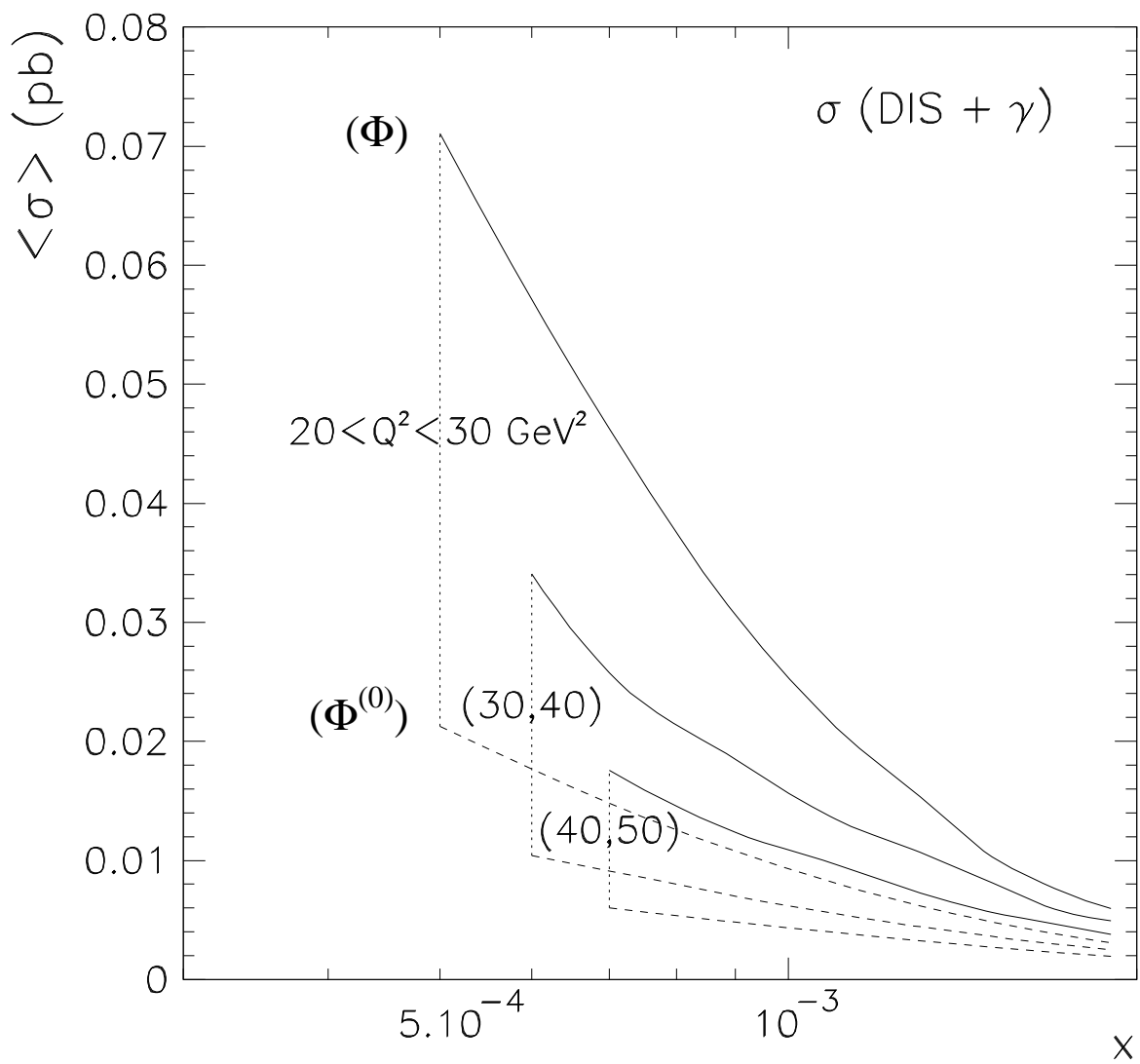
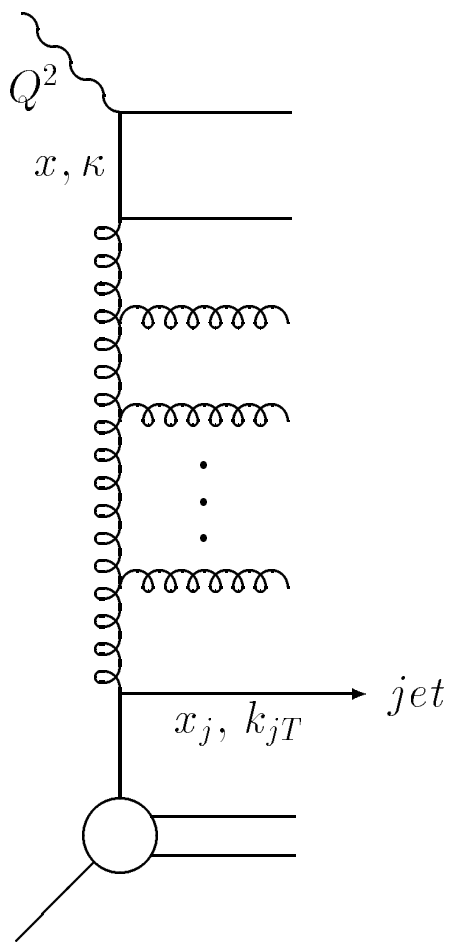
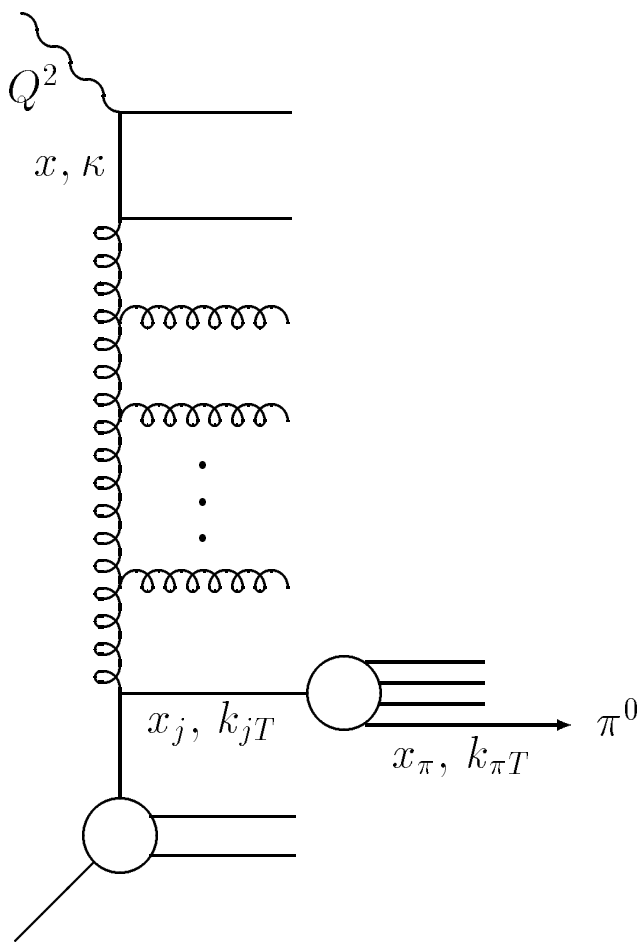
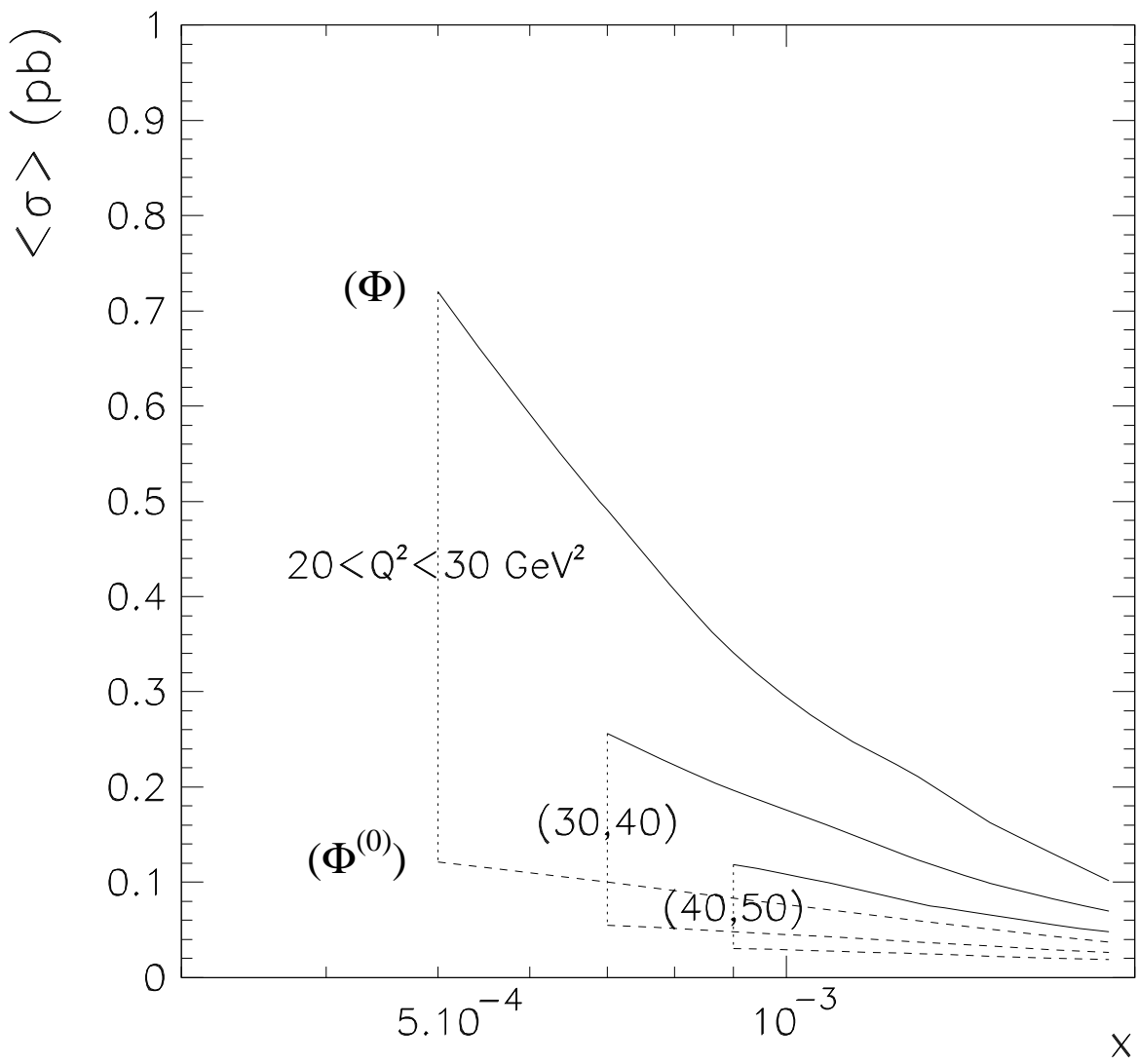
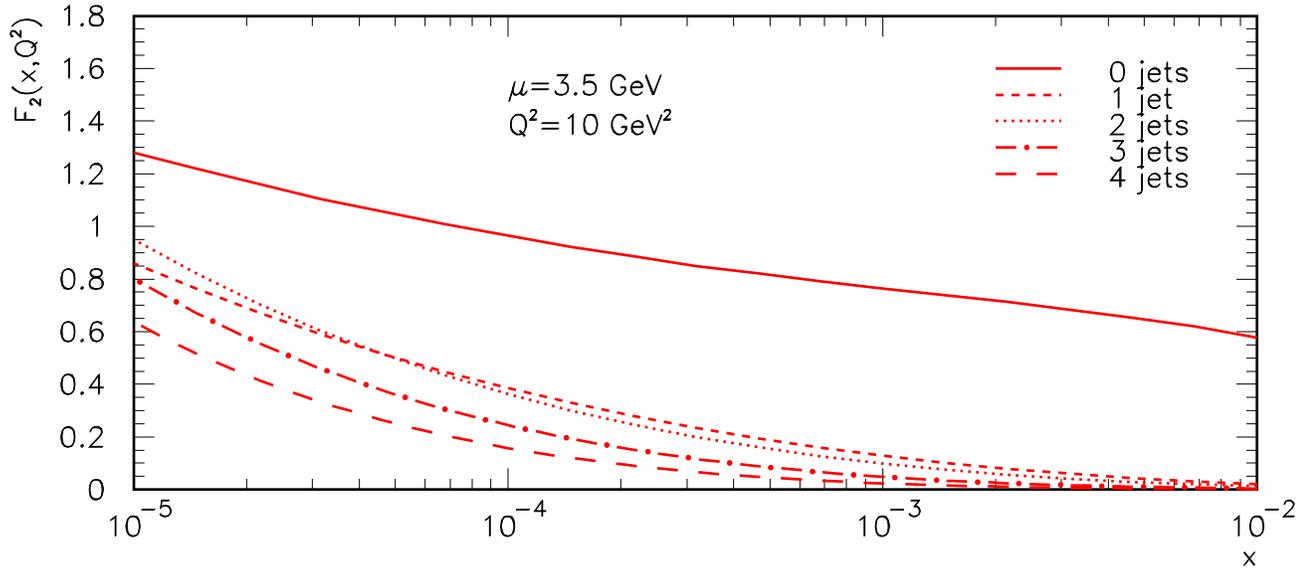
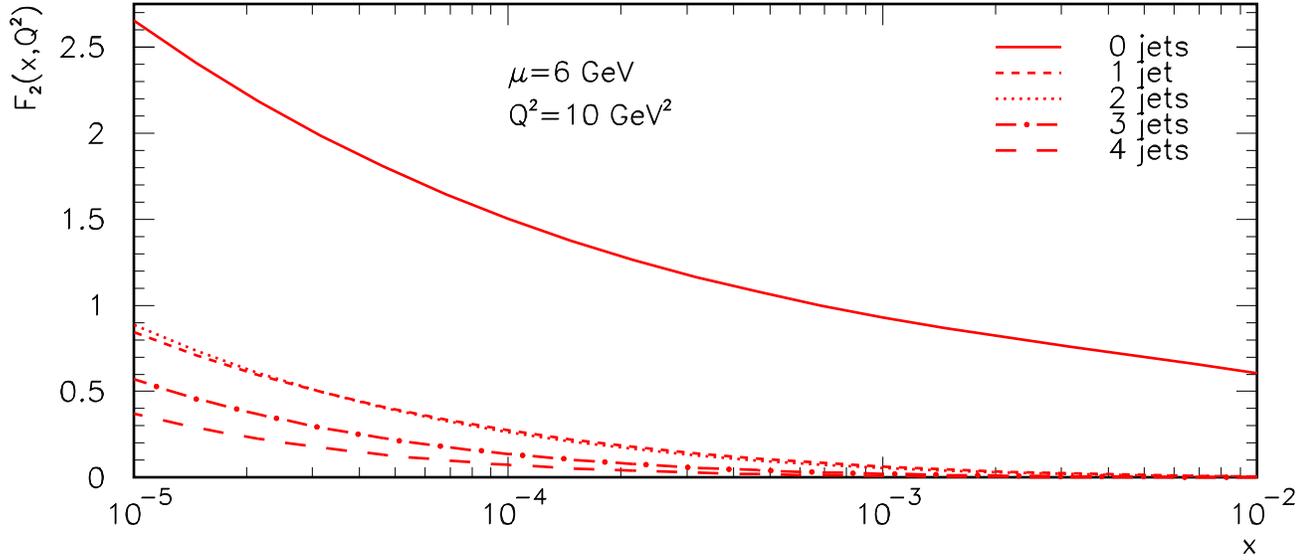
\end{document}